# Giant optical anisotropy in CrSBr from giant exciton oscillator strength

*Georgy Ermolaev[1†], Tagir Mazitov[1†], Arslan Mazitov[1†], Adilet Toksumakov[1†], Dmitriy Grudinin[1†], Anton Minnekhanov[1], Gleb Tselikov[1], Dmitry Yakubovsky[1], Gleb Tikhonowski[1], Nikolay Pak[1], Umer Ahsan[2], Aleksandr Slavich[1], Mikhail Mironov[1], Alexey P. Tsapenko[1], Andrey Vyshnevyy[1], Ivan Kruglov[1], Zdenek Sofer[2], Aleksey Arsenin[1], Kostya S. Novoselov[3,4,5], Andrey Katanin[6], Valentyn Volkov[1\**]*

[1]Emerging Technologies Research Center, XPANCEO, Internet City, Emmay Tower, Dubai, United Arab Emirates

[2]Department of Inorganic Chemistry, University of Chemistry and Technology Prague, Technická 5, 166 28 Prague 6, Czech Republic

[3]The University of Manchester, National Graphene Institute, Oxford Rd, Manchester M13 9PL, U.K.

[4]Department of Materials Science and Engineering, National University of Singapore, Singapore 03-09 EA, Singapore

[5]Institute for Functional Intelligent Materials, National University of Singapore, Building S9, 4 Science Drive 2, Singapore 117544, Singapore

[6]Independent researcher, Moscow, Russia

*Correspondence should be addressed to the e-mail: vsv@xpanceo.com*

## Abstract

**The interplay between dimensionality and electronic correlations in van der Waals (vdW) materials offers a powerful toolkit for engineering light-matter interactions at the nanoscale. Excitons—bound electron-hole pairs—are central to this endeavor, yet maximizing their oscillator strength, which dictates the interaction cross-section, remains a challenge. Conventional wisdom suggests a trade-off, where the observable oscillator strength often decreases in strongly bound systems due to population dynamics. Here, we unveil a colossal oscillator strength associated with the quasi-one-dimensional (quasi-1D) excitons in the layered magnetic semiconductor CrSBr, which fundamentally defies this established scaling law. Through comprehensive optical characterization and ab initio calculations, we establish that this anomalous enhancement originates directly from the reduced dimensionality, which enforces an increased electron-hole wavefunction overlap. Moreover, we find a close connection between fundamental exciton and local spin fluctuations that contribute to the opening of the gap in the electronic spectrum. The resulting optical anisotropy shows a giant in-plane birefringence ($\Delta n \approx$ 1.45) and profoundly anisotropic waveguiding, which we directly visualize using nano-optical imaging. Leveraging this extreme response, we realize a true zero-order quarter-wave plate with an unprecedented wavelength-to-thickness ratio ($\lambda/t$) exceeding 3.4, surpassing the limits of current miniaturization technologies, including state-of-the-art metasurfaces. Our findings underscore the profound impact of dimensionality engineering in magnetic vdW materials for realizing novel regimes of light-matter coupling and developing next-generation ultracompact photonic architectures.**

## Introduction

The excitons, a bound state of an electron-hole pair, stand as fundamental quasiparticles governing the light-matter interaction in semiconductors[1]. Its behavior dictates a vast range of phenomena, from light emission[2] and absorption[3] to energy transfer[4] and the generation of free charge carriers[5]. The properties of excitons, particularly their binding energy and oscillator strength, are therefore of paramount importance in the design and function of virtually all optoelectronic devices, including light-emitting diodes[6], solar cells[7], lasers[8], and photodetectors[9]. The advent of two-dimensional (2D) materials has ushered in a new era of exciton physics, where quantum confinement and reduced dielectric screening give rise to excitons with giant binding energies, ensuring their stability even at room temperature[10]. This robustness has positioned the engineering of excitonic properties in 2D materials at the forefront of materials science, offering a powerful tool to manipulate light at the nanoscale[11].

A particularly compelling manifestation of exciton resonances is optical anisotropy[12,13]. This property, quantified by birefringence $\Delta n$, arises from asymmetries in a material's crystal structure and atomic bonding[14–17]. In excitonic van der Waals (vdW) materials, this optical anisotropy is amplified to an extraordinary degree[12,18–25] of up to 3.0 for out-of-plane vdW direction and up to 2.0 for in-plane vdW direction. With the introduction of 2D magnets, such as CrSBr[26,27], NiPS$_3$[1,28], and CrCl$_3$[29,30], a new dimension has opened up to control excitonic properties through coupling to magnetic order.

Among these materials, CrSBr stands out as a unique platform in which structural, electronic, and magnetic anisotropies converge to create exceptional optical phenomena[26,31,32]. CrSBr is an A-type antiferromagnet (AFM) semiconductor with a high Neel temperature ($T_N \approx 132$ K) and remarkable stability under ambient conditions, which distinguishes it from other 2D magnets[31,32]. The fundamental origin of its anisotropy is the orthorhombic crystal structure (space group *Pmmn*), which gives its electronic zones a quasi-one-dimensional (quasi-1D) behavior[33]. First-principles calculations reveal a nearly flat conduction zone along the crystallographic *a*-axis and a strongly dispersing zone along the *b*-axis, resulting in the effective electron mass ratio reaching 50[10,33]. As a consequence, excitons in CrSBr inherit this quasi-one-dimensional (quasi-1D) nature, forming strongly bound states with wavefunctions spatially stretched predominantly along the *b*-axis[34]. This anisotropy results in giant linear dichroism, where the fundamental exciton transition at ~1.35 eV is intensely bright for light polarized along the *b*-axis, but optically forbidden along the *a*-axis.

The magnetic origin of this exciton in CrSBr leads to numerous phenomena, including magnetically tunable interlayer excitons[10], exciton-magnon interaction[35,36], and the recent discovery of magnetically localized surface excitons[26] and hyperbolic exciton-polaritons[37]. The existence of these effects depends on the exciton resonance with exceptionally large oscillator strength[38,39]. Despite the pivotal role of exciton oscillator strength in governing light-matter interactions in CrSBr, its precise value and direct quantitative link to the material's giant optical anisotropy and peculiarities of magnetic properties have remained elusive.

## Results

### From Crystal Asymmetry to Anisotropic Excitons

The fundamental origin of CrSBr extraordinary optical anisotropy is its profound in-plane and out-of-plane structural anisotropy, seen in Figure 1a and characterized by distinct lattice parameters: *a* = 0.351 nm, *b* = 0.476 nm, and *c* = 0.796 nm[40]. This intrinsic crystallographic asymmetry directly manifests in the material's anisotropy. For example, polarized Raman spectroscopy in Figure 1b reveals a highly anisotropic response for the main $A_g$ phonon modes with a distinct two-lobed scattering pattern. More importantly, this anisotropy governs the quasi-1D nature of the electronic band structure[26,33] resulting in the fundamental exciton with electron-hole wavefunction spatially elongated predominantly along the crystallographic *b*-axis (Figure 1c). A profound impact of the exciton's anisotropy is immediately evident in CrSBr linear optical response. Parallel-polarized optical reflection microscopy in Figure 1d reveals a stark visual contrast dependency on the incident light polarization. CrSBr flake appears intensely bright for light polarized along the crystallographic *b*-axis, corresponding to strong reflection from the exciton. Conversely, it becomes nearly transparent for polarization along the crystallographic *a*-axis, confirming that the exciton transition is optically forbidden along this direction. The resulting giant linear dichroism, originating from CrSBr intrinsic structural asymmetry (Figure 1a), points to an exceptionally large oscillator strength associated with the fundamental exciton, which we proceed to quantify.

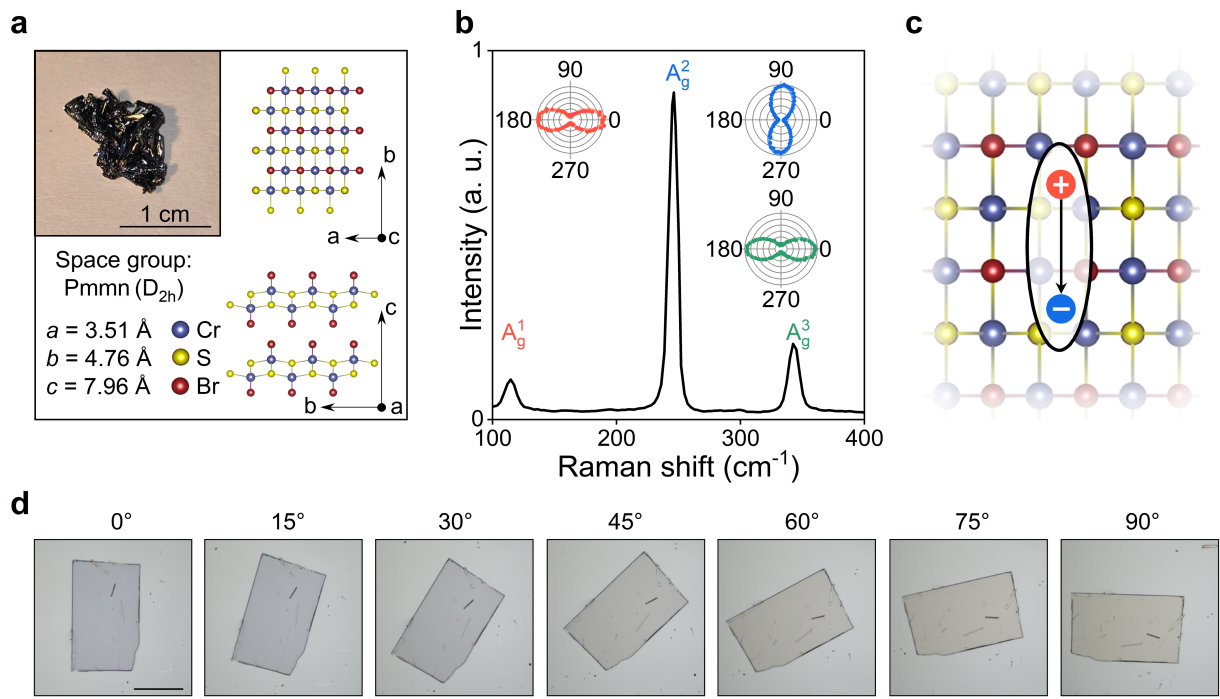

**Figure 1. In-plane structural anisotropy of CrSBr. a,** Orthorhombic crystal structure of CrSBr. The inset shows an optical image of the bulk crystal. **b,** The averaged over polarization angles Raman spectrum of CrSBr for 532 nm excitation laser wavelengths. The insets demonstrate polarized measurements for the main Raman modes: $A_g^1$ (≈115 cm$^{-1}$, red line), $A_g^2$ (≈246 cm$^{-1}$, green line), and $A_g^3$ (≈342 cm$^{-1}$, blue line). **c,** The depiction of the fundamental CrSBr exciton. The solid line represents the binding between an electron and a hole in an exciton. The arrow shows the preferential direction of the exciton along the crystallographic *b*-axis. **d,** Parallel-polarized optical reflection images of CrSBr flake, varying with the rotation angle: 0° and 90° correspond to polarization along the crystallographic *b*-axis and *a*-axis, respectively.

**Giant oscillator strength and anisotropy of 1D exciton**

To quantify the profound optical anisotropy suggested by our reflection microscopy in Figure 2d, we determine the complete refractive index tensor ($n_a + ik_a$, $n_b + ik_b$, $n_c$) of CrSBr (see Methods). The resulting broadband optical constants – the refractive index ($n$) and extinction coefficient ($k$) – are presented in Figure 2a. For light polarized along the crystallographic *b*-axis, the optical response is dominated by an intense exciton resonance centered at 1.38 eV (898 nm). At this energy, the extinction coefficient $k_b$ reaches a peak value of ≈1.8, indicating exceptionally strong light absorption. In stark contrast, the optical response for light polarized along the *a*-axis is nearly featureless across the same spectral range, with a negligible extinction coefficient $k_a \approx 0$. This observation confirms that the fundamental exciton transition is optically allowed only for *b*-axis polarization and is forbidden along the *a*-axis[34], giving rise to giant linear dichroism. As a result, the strong resonance in $k_b$ induces a large dispersion in $n_b$ via the Kramers-Kronig relations, leading to a substantial in-plane birefringence, $\Delta n = n_b - n_c$ (Figure 1a).

These experimental findings are in excellent agreement with our ab initio many-body perturbation theory calculations (see Methods), presented in Figure 2b. The calculations, performed using the GW approximation and the Bethe-Salpeter equation (GW-BSE), quantitatively reproduce the spectral position, asymmetric lineshape, and polarization direction of the fundamental exciton. The excellent correspondence between theory and experiment confirms that the observed giant optical anisotropy is an intrinsic property of CrSBr, stemming directly from its unique electronic and crystal structure.

The pronounced excitonic peak is a direct manifestation of an exceptionally large oscillator strength, determined by the degree of the overlap between the electron and hole of exciton wavefunctions and exciton population[42]. The former shows the intrinsic oscillator strength of a single exciton, which strongly depends on the exciton binding energy ($E_b$) – the energy required to separate the electron and hole into free carriers[3,29]. A larger $E_b$ corresponds to a smaller exciton Bohr radius ($a_b$), resulting in larger electron and hole wavefunctions overlap, i.e., a larger intrinsic oscillator strength. This relation explains why materials with large $E_b$ are not only thermally stable but are also intrinsically brighter optical emitters with strong light-matter interaction[43]. However, if we compare the exciton oscillator strength (the maximum value of the imaginary part of the exciton dielectric function $\varepsilon$, max[Im($\varepsilon$)]) for traditional vdW materials in Figure 2c, one can notice the reverse behavior – increasing the binding energy decreases the exciton oscillator strength. The primary reason for this trend is a latter factor of exciton population, governed by the universal Arrhenius law of exponential dependence on the activation energy, which, in the exciton case, is a binding energy[29]. Indeed, it perfectly describes the relation between the exciton oscillator strength and the binding energy in Figure 2c. In this regard, CrSBr occupies a unique position in Figure 2c, exhibiting a giant oscillator strength for a given binding energy, far exceeding that of conventional vdW materials. Since the exciton population follows the Arrhenius law, the only origin of this anomaly is an extremely large intrinsic oscillator strength, determined by the overlap of the electron and hole wavefunctions. Therefore, the giant exciton oscillator strength in CrSBr is a direct consequence of the exciton quasi-1D nature because the volume occupied by the exciton scales approximately as $a_b^3$ in 3D, $a_b^2$ in 2D, and $a_b$ in 1D, implying larger values for the reduced dimensionality. Thus, Figure 2c validates the 1D character of the fundamental exciton and its great potential in excitonic applications.

For example, this giant anisotropic oscillator strength results in a colossal in-plane birefringence of CrSBr, which we plot in Figure 2d. The birefringence reaches a peak value of $\Delta n \approx 1.45$ near the exciton resonance and maintains a remarkably high value of $\Delta n$ from 0.25 to 0.52 in the material's transparency region ($\lambda$ > 1100 nm). As seen in Figure 2d, this optical anisotropy is the highest among other anisotropic materials transparent in the near-infrared spectral range. As a result, it establishes CrSBr as a premier platform for ultracompact nanophotonic devices, enabling functionalities such as polarization control and anisotropic waveguiding at the nanoscale.

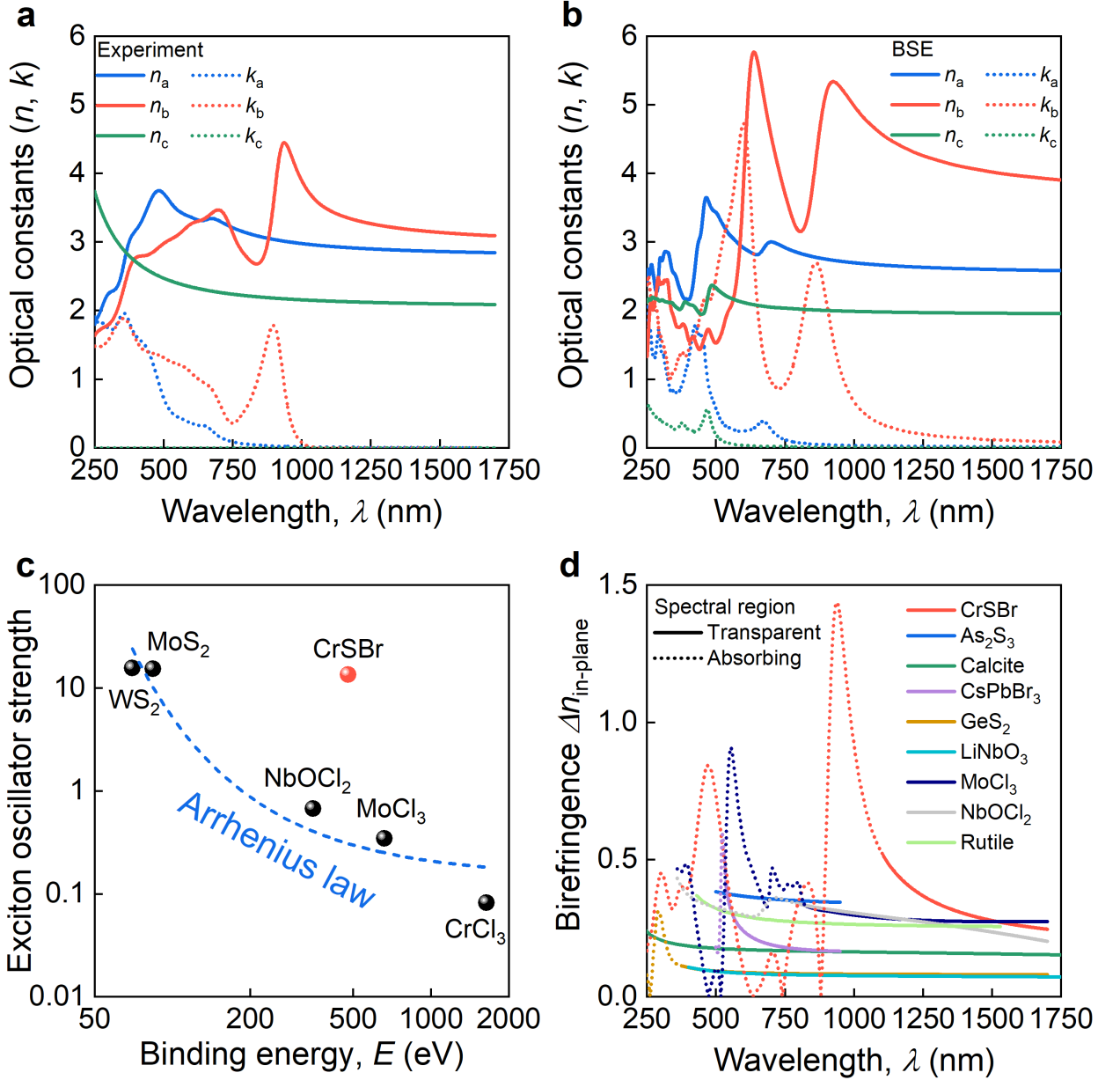

**Figure 2. In-plane optical anisotropy of CrSBr. a,** Broadband anisotropic optical constants of CrSBr. **b,** Ab initio calculations of anisotropic optical constants of CrSBr. **c,** The dependence of the exciton oscillator strength in vdW materials on the exciton binding energy. CrSBr demonstrates the giant exciton oscillator strength for a given binding energy. **d,** The comparison of in-plane birefringence among transparent in near-IR anisotropic semiconductors. CrSBr demonstrates the record value in the transparency region. The optical constants for other materials are adopted from several works[13,17,29,44,45].

**Microscopic origin of exciton properties: the results of many-body approaches**

To unravel the microscopic origins of the colossal oscillator strength, we first turn to ab initio many-body perturbation theory, whose predictions for the optical constants (Figure 2b) show good agreement with our experimental findings (Figure 2a). While these calculations, performed using the Bethe-Salpeter equation on top of a self-consistent GW foundation (GW@scGW), assume a magnetically ordered A-type antiferromagnetic ground state ($T$ = 0 K), they remarkably capture the spectral position of the fundamental exciton observed at room temperature (Figures 2a-b). This correspondence suggests that the key electronic features governing the exciton are robust against thermal fluctuations. To reconcile the quantitative differences in optical constants' magnitude and validate this picture, we employed a combination of density functional theory and dynamical mean-field theory (DFT+DMFT) to model the electronic properties in the high-temperature paramagnetic phase. The results, presented in Figure 3a, provide a decisive justification. The GW band structure of the ordered phase (black lines) aligns perfectly with the regions of high spectral density of Cr d-orbitals in the paramagnetic phase (color map). Moreover, while analyzing the contributions to oscillator strength of the fundamental exciton at high symmetry points, we noticed that they reach the highest values at the gap boundaries, which are located in the regions of high spectral density of d-orbitals in the paramagnetic state at room temperature. This confirms that the electronic gap and the character of the band-edge states are governed by on-site Coulomb interactions and Hund's coupling that produce robust local magnetic moments, rather than the long-range magnetic order. More detailed analysis (see Supplementary Note 1) shows that the Hund coupling plays the decisive role, stressing therefore the importance of local spin fluctuations for opening the gap in the electronic spectrum. From this perspective, the suppression of the refractive index and extinction coefficient in the experimental data compared to the T = 0 K theory is naturally explained: the notable broadening of the spectral density in the paramagnetic state (Figure 3a) reflects a reduced quasiparticle lifetime due to incoherent scattering on local spin fluctuations, which inherently dampens the optical constants. Nonetheless, the optical response of these interactions is still sufficient at room temperature to demonstrate the record optical characteristics (Figures 2c-d).

In addition to the robustness of the electronic structure across the magnetic phase transition, we now demonstrate that it has a profound spatial anisotropy, the quality that defines the quasi-1D exciton (Figure 3b). A similar anisotropy is obtained for magnetic exchange interactions in Figure 3c, which reveals a highly anisotropic magnetic landscape. Along the crystallographic *b*-axis, the interactions are strongly ferromagnetic (yellow circles) and decay weakly with distance, forming a continuous pathway for magnetic coupling. In stark contrast, interactions along the *a*- and *c*-axes are significantly weaker and decay rapidly. This anisotropic network of exchange interactions, which is mediated by the overlap of electron orbitals, reflects a quasi-1D character of electron and hole motion. The exciton, as a bound electron-hole pair, minimizes its energy by delocalizing along this path of strongest electronic coupling. This is visualized by comparing the magnetic interaction map with the calculated real-space distribution of the exciton wavefunction in Figure 3b. The wavefunction is elongated along the *b*-axis (Figure 3b), perfectly mirroring the spatial distribution of the exchange pathways (Figure 3c).

This confirms that the quasi-1D behavior of electron wavefunctions is not a fragile, low-temperature phenomenon but an intrinsic property of CrSBr that persists up to and beyond room temperature. We also demonstrate this effect through calculations of the spin stiffness (Figure 3d), a microscopic measure of the energy required to create a spin excitation, which reflects the collective strength and dimensionality of the magnetic order. The calculated spin stiffness surface at room temperature (T = 290 K), plotted in Figure 3d, is clearly elongated along the *b*-axis. The evolution of the spin stiffness with temperature, shown in Figure 3e, reveals that this profound anisotropy is already well-established at high temperatures and becomes stronger upon cooling. Therefore, this quasi-1D shape of the magnetic response, which reflects anisotropy of electronic properties, is an intrinsic, high-temperature feature of the material, not an emergent property that appears only near Neel temperature $T_N \approx 132$ K.

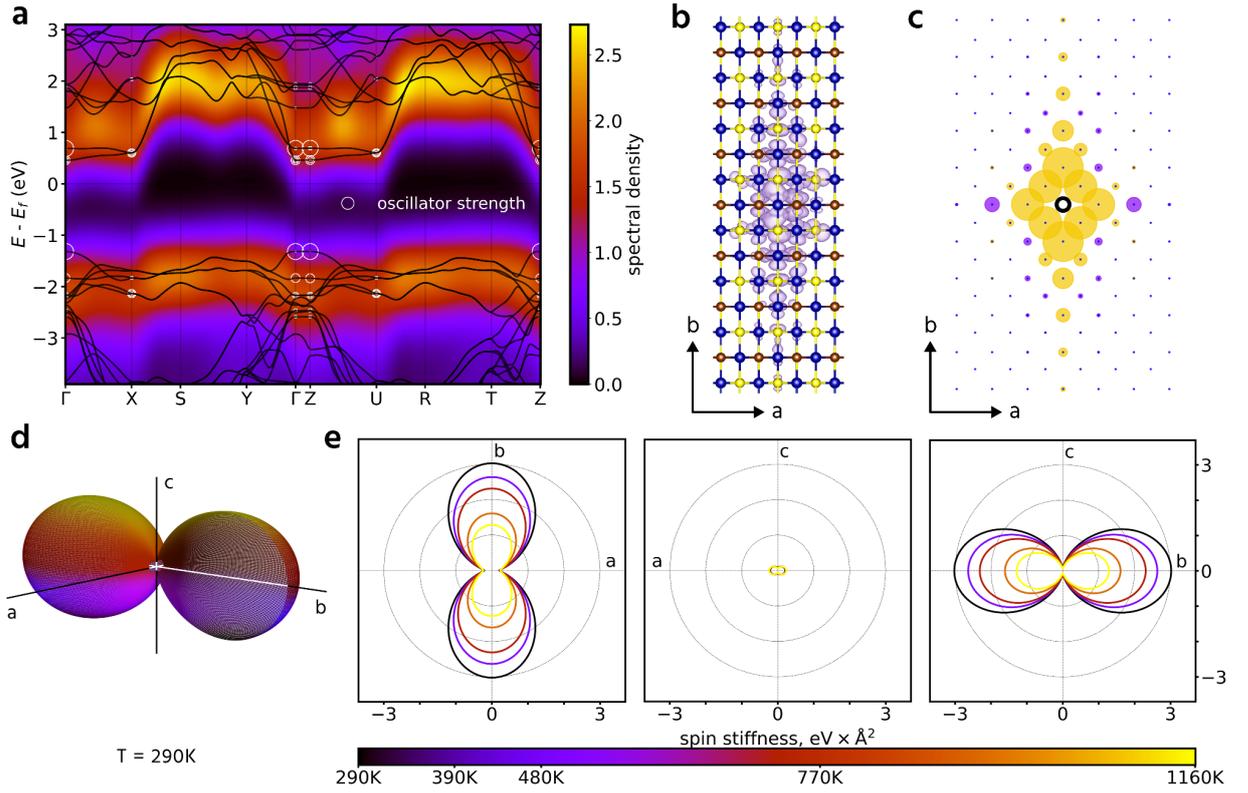

**Figure 3. Theory behind the giant oscillator strength of CrSBr. a,** Band structure in the GW approach (black lines) and exciton oscillator strength at high-symmetry points (white circles), compared to the spectral density of d-states in the DFT+DMFT approach (color scale). **b,c,** Real space distribution in a-b plane of wave functions of **b,** electrons and holes in exciton and **c,** exchange interactions (ferromagnetic interaction is shown by yellow circles, antiferromagnetic by purple circles; sizes of circles demonstrate the value of interaction with the central atom). **d,** 3D-surface of spin stiffness in various directions of momentum space at room temperature (290K), *a,b,c* being crystallographic axes. **e,** 3D-surface cuts of spin stiffness along *a-b*, *a-c*, *b-c* crystallographic planes with temperature evolution shown in color (see the colorbar below).

**Direct visualization of anisotropic waveguiding**

To explore the quasi-1D characteristics of the in-plane optical anisotropy and guided mode behavior in CrSBr, we perform scattering-type scanning near-field optical microscopy (s-SNOM) in the reflection configuration, shown in Figures 4a-d (see Methods for the experimental details). Two orthogonal sample orientations, labelled XY and YX (Figures 4a-d), were investigated by rotating the flake by 90° with respect to the laser propagation direction. This geometry allows us to individually visualize waveguide properties along the crystallographic *a*-axis and *b*-axis.

Near-field amplitude and phase s-SNOM images in Figures 4e and 4g show high-contrast interference fringes of guided modes in CrSBr. Their Fast Fourier transforms (FFTs) in Figures 4f and 4h reveal multiple discrete momentum peaks corresponding to guided modes in CrSBr. To quantify the observed modes, we compare their experimental momenta from Figures 4f and 4h with the theoretical mode dispersion in Figures 4i-j calculated using the transfer-matrix method[46] for Air/CrSBr (295 nm)/SiO$_2$ (275 nm)/Si heterostructure. Notably, for the excitation wavelength 900 nm corresponding to the exciton, CrSBr supports only TM$_0$ mode along the crystallographic *a*-axis (Figure 4i) because the electric field strongly interacts with the exciton, which supresses the TE$_0$ mode. Conversely, along the crystallographic *b*-axis, CrSBr supports all guiding modes (Figure 4h) since the electric field interaction with the exciton is minimal for them. For larger wavelengths, exciton absorption drops, and all guided modes propagate inside CrSBr, as shown in Figures 4i-j and Supplementary Note 2. Nevertheless, we can clearly see the tremendous difference in the waveguide in-plane momenta along the crystallographic *a*-axis and *b*-axis (Figures 4i-j). Altogether, these findings provide direct nanoscale evidence of anisotropic waveguiding in CrSBr, demonstrating a perfect agreement between experiment and theory and reinforcing the anisotropic optical constants in Figure 2a.

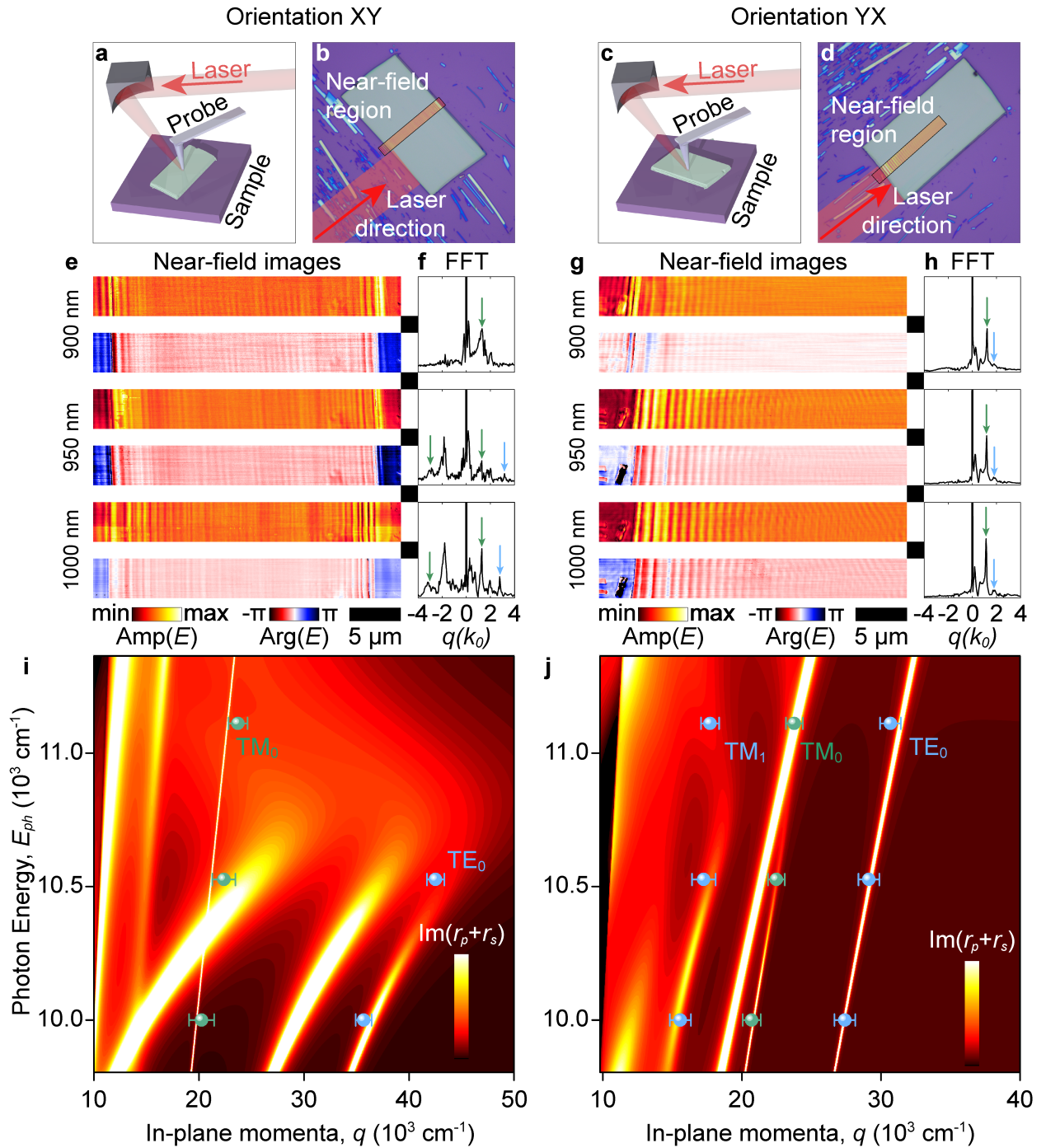

**Figure 4. Anisotropic waveguiding in CrSBr. a,** The near-field scheme for recording CrSBr waveguiding properties along the crystallographic *a*-axis and **b,** optical image of CrSBr flake in this setup. **c,** The near-field scheme for recording CrSBr waveguiding properties along the crystallographic *b*-axis and **d,** optical image of CrSBr flake in this setup. **e,** Near-field images: amplitude Amp(*E*) and phase Arg(*E*) of the waveguide and **f,** their Fourier transform amplitude for the configuration in panels (**a**) and (**b**). **g,** Near-field images: amplitude Amp(E) and phase Arg(E) of the waveguide and **h,** their Fourier transform amplitude for the configuration in panels (**c**) and (**d**). Transfer-matrix calculations for propagating modes inside CrSBr along the crystallographic **i,** *a*-axis and **j,** *b*-axis.

**An ultracompact wave plate with record-breaking performance**

In addition to anisotropic waveguiding (Figure 4), the giant optical anisotropy of CrSBr in Figure 2d positions it as a promising platform for the realization of ultracompact nanophotonic components for polarization control. To demonstrate this potential, we harness CrSBr birefringence to fabricate a true zero-order quarter-wave plate (QWP), a fundamental optical element that converts linearly polarized light into circularly polarized light (Figure 5a). The device concept and experimental configuration are illustrated in Figures 5a-b, a CrSBr flake of thickness $t$ = 305 nm, exfoliated onto a transparent $CaF_2$ substrate, serves as QWP. We first characterize the device's performance by measuring the phase retardance on transmitted light as a function of wavelength (Figure 5c, see Methods for the experimental details). The device achieves a phase retardance of 90°, the essential requirement for QWP operation, at two distinct wavelengths in the near-infrared: 940 nm and 1029 nm. This dual-wavelength functionality is a hallmark of operating with a high optical anisotropy[17]. In conventional wave plates, the retardance $\delta$ is described by the simple relation $\delta = (2\pi \Delta n t)/\lambda$. However, for a material with high refractive index and high optical anisotropy, such as CrSBr, the retardance acquires an additional factor of Fabry-Perot resonances within the nanoscale-thick flake, creating a hybrid operation mode that combines classical birefringent phase delay with resonant cavity effects[17].

To validate the QWP functionality, we performed angle-resolved transmittance measurements, the results of which are presented as maps of transmitted intensity versus polarizer ($\theta_P$) and analyzer ($\theta_A$) angles in Figures 5d-e. The definitive signature of the conversion from linear to circular polarization is that the transmitted intensity becomes independent of the analyzer angle $\theta_A$. At the operational wavelength of 1029 nm (Figure 5e), this condition is met when the input polarizer is set to $\theta_P$ = 44.0°, resulting in a nearly uniform vertical line on the transmittance map. At 940 nm (Figure 5d), however, optimal conversion to circular polarization occurs at a polarizer angle of $\theta_P$ = 68.5°. This deviation from the classical $\theta_P$ = 45° is another direct manifestation of CrSBr extreme anisotropy because, for ideal QWP operation, the incident linear polarization must be aligned at an angle $\theta_P$ = arctan($|t_b|/|t_a|$), where $|t_b|$ and $|t_a|$ are the transmission amplitudes along the crystal's principal optical axes. Due to the giant oscillator strength (Figure 2), the exciton resonance renders these transmission amplitudes unequal ($|t_b|$ ≠ $|t_a|$), thus shifting the optimal input angle away from 45°. This result provides a direct link between the functional performance of the device and the fundamental electronic properties of CrSBr.

Finally, to establish the significance of our device in the context of optical miniaturization, we benchmark its performance against other state-of-the-art ultrathin QWPs (Figure 5f). The key figure of merit for a miniaturized wave plate is the ratio of its operating wavelength $\lambda$ to its thickness $t$, which quantifies its compactness. Our CrSBr QWP achieves an exceptional $\lambda/t$ ratio of ≈3.4 at 940 nm and ≈3.1 at 1029 nm. As shown in Figure 5f, this performance surpasses that of wave plates based on other advanced platforms, including $As_2S_3$[17], $Fe(C_5H_5)_2$[47], polymer gratings[48], and even highly optimized dielectric metasurfaces[49]. The CrSBr device in Figure 4 represents an unprecedented leap in miniaturization, more than doubling the performance of the previous record-holders[17]. This result unequivocally establishes CrSBr as a perspective material for ultracompact, high-performance nanophotonic devices, a direct consequence of its fundamental feature of quasi-1D exciton.

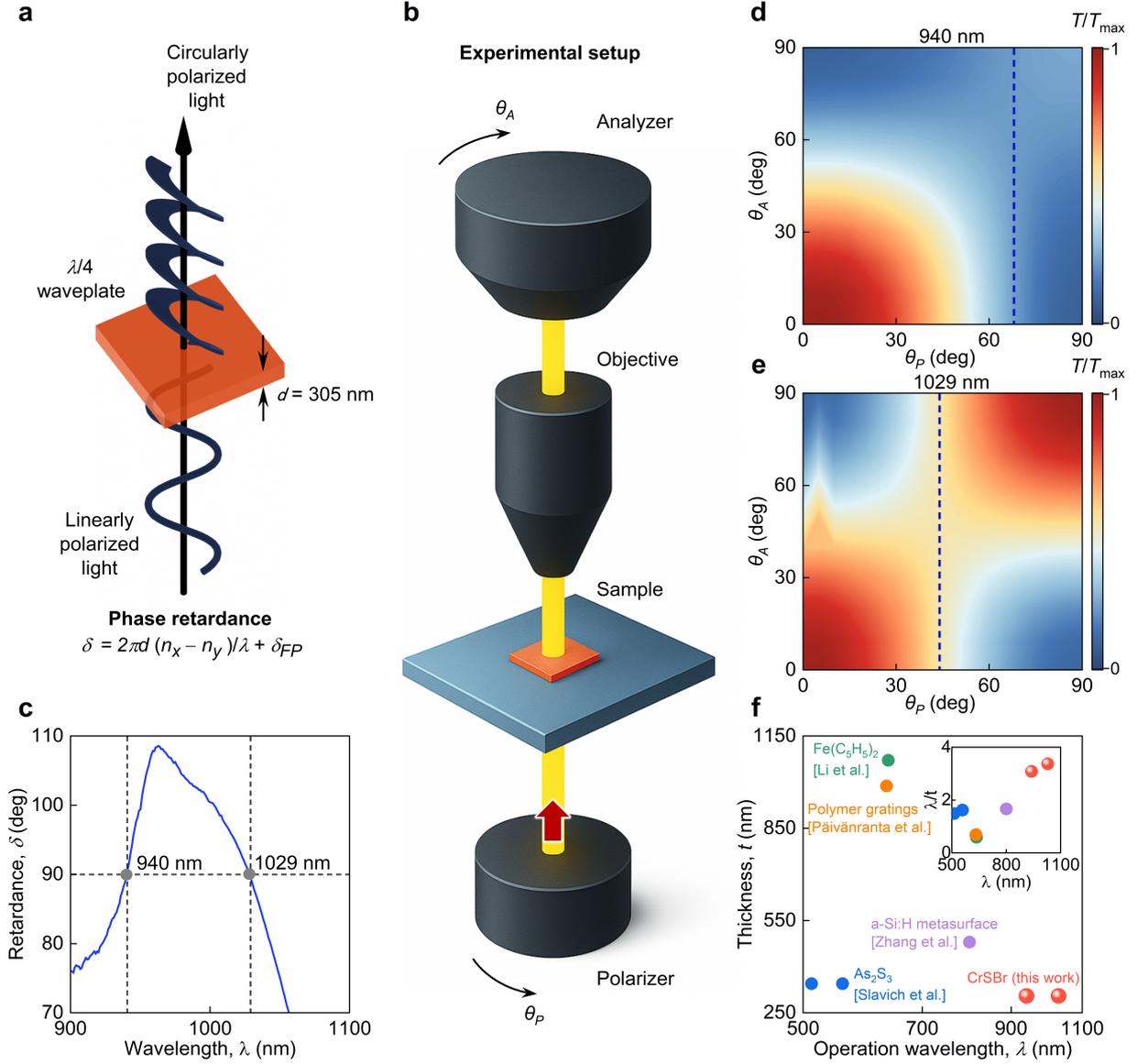

**Figure 5. CrSBr record quarter-wave plate. a,** The scheme of CrSBr quarter-wave plate. **b,** Schematic representation of the experimental setup. **c,** The dependence of phase retardance on wavelength for the CrSBr wave plate. The dashed lines show the quarter-wave operation regime (retardance = 90°). The transmittance map of the CrSBr waveplate dependence on analyzer $\theta_A$ and polarizer $\theta_P$ angles for **d,** 940 nm and **e,** 1029 nm. The dashed lines show the quarter-wave operation regime, when linear polarization transforms to circular polarization: $\theta_P$ = 68.5° for 940 nm and 44.0° for 1029 nm. **f,** The comparison of CrSBr wave plate operations with other ultrathin wave plates from other works: Fe(C$_5$H$_5$)$_2$[47], polymer gratings[48], a-Si:H metasurface[49], and As$_2$S$_3$[17]. The inset shows the comparison of the ratio between the operation wavelength and the thickness of the wave plate.

## Conclusions

In conclusion, we have elucidated the fundamental origin of the giant optical anisotropy in vdW magnetic semiconductor CrSBr, tracing it directly to the colossal oscillator strength of its intrinsic quasi-1D exciton[10,26,33]. This dimensional reduction dramatically enhances the electron-hole wavefunction overlap, resulting in a colossal oscillator strength that is anomalous when compared to the established trend in other van der Waals materials, where oscillator strength typically diminishes with increasing exciton binding energy. The technological relevance of this fundamental discovery is demonstrated through the direct visualization of highly anisotropic waveguiding in CrSBr and the realization of an ultracompact, true zero-order quarter-wave plate (QWP) that operates at the ultimate limits of miniaturization. As a result, our work demonstrates that a top-down approach, centered on the discovery and application of materials with extreme intrinsic properties[14,16,17], can offer a powerful and simple path to radical miniaturization. This result challenges the prevailing view that metasurfaces are the sole path forward for ultracompact optics[50,51] and reestablishes the central role of materials discovery in advancing nanophotonics.

## Acknowledgement

Z.S. was supported by project LUAUS25268 from Ministry of Education Youth and Sports (MEYS) and by the project Advanced Functional Nanorobots (reg. No. CZ.02.1.01/0.0/0.0/15_003/0000444 financed by the EFRR).